\documentclass[aps,prb,reprint,amssymb,amsmath,superscriptaddress]{revtex4-1}

\usepackage{graphicx}
\usepackage{epstopdf}
\usepackage[english]{babel}
\bibpunct{[}{]}{,}{n}{}{}

\begin{document}

\title{Spin-orbit splitting of valence and conduction bands in HgTe quantum wells near the Dirac point}

\author{G.\,M.~Minkov}

\affiliation{Institute of Natural Sciences, Ural Federal University,
620002 Ekaterinburg, Russia}

\author{A.\,V.~Germanenko}

\author{O.\,E.~Rut}
\affiliation{Institute of Natural Sciences, Ural Federal University,
620002 Ekaterinburg, Russia}

\author{A.\,A.~Sherstobitov}

\affiliation{Institute of Natural Sciences, Ural Federal University,
620002 Ekaterinburg, Russia}

\affiliation{M.~N.~Miheev Institute of Metal Physics of Ural Branch of
Russian Academy of Sciences, 620137 Ekaterinburg, Russia}

\author{M.\,O.~Nestoklon}

\affiliation{Ioffe Physical-Technical Institute, Russian Academy of Sciences, 194021 St. Petersburg, Russia}

\author{S.\,A.~Dvoretski}

\affiliation{Institute of Semiconductor Physics RAS, 630090
Novosibirsk, Russia}

\author{N.\,N.~Mikhailov}

\affiliation{Institute of Semiconductor Physics RAS, 630090
Novosibirsk, Russia}

\date{\today}

\begin{abstract}
Energy spectra both of the conduction and valence bands of the HgTe quantum wells with a width close to the Dirac point were studied experimentally. Simultaneous analysis of  the Shubnikov-de Haas oscillations and Hall effect over a wide range of electron and hole densities gives surprising result: the top of the valence band is strongly split by spin-orbit interaction while the splitting of the conduction band is absent, within experimental accuracy.  Astonishingly, but such a ratio of the splitting values is observed as for structures with normal spectrum so for structures with inverted one. These results do not consistent with the results of \emph{kP} calculations, in which the smooth electric filed across the quantum well is only reckoned in. It is shown that taking into account the asymmetry of the quantum well interfaces within a tight-binding method gives reasonable  agreement with the experimental data.
\end{abstract}


\maketitle

\section{Introduction}
\label{sec:intr}
Heightened interest to two-dimensional (2D) structures with quantum well of gapless semiconductors is caused by the fact that different types of the carrier energy spectrum could be realized in these structures depending on the width ($d$) of quantum well. The most detailed, both theoretically and experimentally, heterostructures with Hg$_{1-x}$Cd$_x$Te/HgTe/Hg$_{1-x}$Cd$_x$Te quantum well were studied. The calculations of the energy spectrum in framework of \emph{kP} method for symmetrical quantum well show that there is the critical  width of the HgTe quantum well, $d=d_c\simeq 6.3$~nm, when the linear in quasimomentum Dirac-like energy spectrum should be realized at small quasimomentum ($k$) \cite{Gerchikov90,Bernevig06}. At $d<d_c$ the energy spectrum is normal.  The valence band is formed from heavy-hole states while the conduction band is formed from  electron states and states of light hole. At $d>d_c$  these states swap over, and such a spectrum is called an inverted spectrum. Another approach, the calculation in the framework of tight-binding model, which takes into account  the bulk inversion asymmetry  of the zinc blende lattice, gives close result regarding the dependence $E(k)$ except the fact that  the relatively large anti-crossing at $k=0$ and large spin-orbit (SO) splitting of the conduction and valence band appears at $d=d_c$.

Knowledge of the energy spectrum is necessary for an understanding of all properties of 2D systems: optical, transport and others. However it has been experimentally studied  rather superficially to date.  The energy spectrum of the conduction band was studied more detailed \cite{Schultz96,Pfeuffer98,Schultz98,Zhang04,Ikonnikov11}. There was shown that the SO splitting of the conduction band in the technologically symmetric quantum wells   was not reveal itself in the most cases, the effective mass of the electrons increased with their density. These data were found in satisfactory agreement with results of theoretical calculations performed in the  framework of the \emph{kP} method.

The energy spectrum of the valence band is investigated still less and, what is important, the experimental data are inconsistent with theoretical results in many cases. So in the structures with  $d>10$~nm that corresponds to the inverted energy spectrum, the hole effective mass ($m_h$) within wide hole density range $p=(1-4)\times 10^{11}$~cm$^{-2}$ occurs substantially less than that calculated within \emph{kP} method: $m_h\simeq (0.15-0.3)m_0$ \cite{Kozlov11,Kvon11-1,Minkov13} instead of $(0.5-0.6)m_0$ \cite{ZholudevPhD}. These calculations predict also that the conduction band should overlap with the valence band which top is located at $k\neq 0$ when $d\gtrsim (12-15)$~nm. Although this prediction is in qualitative agreement with experimental data \cite{Olshanetsky12, Minkov13}, the quantitative  difference between theory and experiment is drastic.
Experimentally, the top of the valence band is located at $k\simeq 0.5\times 10^6$~cm$^{-1}$, while the theoretical prediction gives the value of about $k\simeq 2.5\times 10^6$~cm$^{-1}$.

The valence band spectrum in the heterostructures with normal band ordering was studied in several papers only. The experimental data published in Refs.~\cite{Ortner02} and~\cite{Minkov14} for structures with $d=(5-6)$~nm are very similar however the interpretation differs significantly. Analyzing the Hall density and the Fourier spectra of the Shubnikov-de Haas (SdH) oscillations, the authors of Ref.~\cite{Ortner02} are able to describe the data taking into account the secondary maxima of the dispersion law located at $k\neq 0$. But it has been done only for one hole density. Studying analogous heterostructures, the authors of Ref.~\cite{Minkov14} show that such interpretation does not describe the experimental data within a wide hole density range. They show that all the results are well described under assumption that the top of the valence band is very strongly split by SO interaction. They supposed that the reason for such a splitting is a strong electric field of \emph{p-n} junction of technological origin in which the quantum is embedded. The authors were unable to investigate the splitting of the conduction band. These data would make the interpretation more reliable.

In the present paper, we report the results of an experimental study both of hole and electron transport in HgTe quantum well of different width near the critical point $d_c$ with a normal and inverted energy spectra. The measurements were performed over a wide range of carrier density. It has been experimentally found that the valence band is strongly split by SO interaction, while the conduction band remains unsplit independently of energy band ordering. We believe that the natural interface inversion asymmetry of zinc-blende heterostructures is responsible for these peculiarities. Quantitatively, the experimental data are well described in the framework of atomistic calculation \cite{Tarasenko15}.

\section{Experimental}
\label{sec:expdet}

Our samples with HgTe quantum wells  were realized on the basis of
HgTe/Hg$_{1-x}$Cd$_{x}$Te ($x=0.55-0.65$) heterostructures grown by
molecular beam epitaxy on GaAs substrate with the (013) surface
orientation \cite{Mikhailov06}.  The samples were mesa etched into standard Hall bars of
$0.5$~mm  width and the distance between the potential probes of
$0.5$~mm. To change and control the carrier density in the quantum
well, the field-effect transistors were fabricated with parylene as an
insulator and aluminium as a gate electrode. For each heterostructure,
several samples were fabricated and studied. The measurements
were performed at temperatures of $1.3-20$~K.

\section{Characterization of samples regarding the type of spectrum}
\label{sec:char}

It is very important for interpretation of the experimental results to know whether the spectrum of structure under study is normal or inverted. The values of quantum well width presented in the Table~\ref{tab1} are technological, therefore it is very desirable to have independent data on the spectrum type.

\begin{figure}
\includegraphics[width=\linewidth,clip=true]{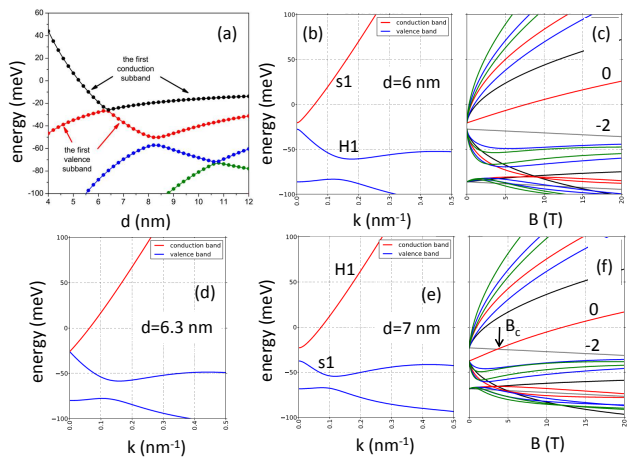}
\caption{(Color online) The spatial quantization subband energies at $k = 0$ plotted against the HgTe quantum well width. The band structure (b), (d), (e) and Landau levels (c), (f) for quantum well of different widths \cite{ZholudevPhD}.  }\label{f1}
\end{figure}

The most reliable method to determine the type of the spectra can be based on the peculiarity of the spectrum quantization in the external magnetic field. Theoretical calculations \cite{Schultz98,Koenig07,ZholudevPhD} and experimental investigations \cite{Koenig07,Minkov13} show that there are two anomalous Landau levels which behavior is radically different for the normal and inverted spectra [Figs.~\ref{f1}(c) and \ref{f1}(f)].

As seen from Fig.~\ref{f1}(f)  the Landau levels $-2$ and  $0$ in the structure with inverted spectrum start at $B=0$  from the bottom of the conduction band and top of the valence, respectively, and moving towards each other cross in the magnetic field $B=B_c$. This magnetic field increases with $d$ increase, achieves  maximal value of  about $9-10$~T at $d\simeq (9-10)$~nm and then decreases with a further $d$ increase.  In the structures with normal spectrum, the energy positions of the anomalous Landau levels at $B\to 0$ is opposite: the level $-2$ starts from the valence band top while the level $0$ goes from the conduction band bottom. Because they move in the same directions as for $d>d_c$,  the crossing does not occur in this case [Fig.~\ref{f1}(c)]. Thus, if  the crossing of Landau levels  is detected experimentally, the real width of the HgTe quantum well is greater than critical value $d_c$ and the sample under study is in inverted regime.

When $B_c$ is large enough so that the anomalous Landau levels  have a large density of states and are well separated,  the cross manifests itself as the nonmonotonic peculiarity in the $\rho_{xx}$~vs~$B$ and $\rho_{xy}$~vs~$B$ dependences \cite{Minkov13}. When the quantum well width is close to the critical value $d_c$, the cross of the Landau levels occurs at so low magnetic field that it may not reveal itself in magnetotransport measurements.  It is possible in this case to observe the cross by studying the behavior of the quantum capacitance ($C_q$) in the magnetic field at the gate voltages close to the charge neutrality point (CNP). Unlike the resistance components which depend not only on the density of states ($\nu$) but on the disorder strength also, the quantum capacitance depends on the density of states  only, $C_q=e^2\, dn/d\mu=e^2\,\nu(\mu)$, where $\mu$ stands for the chemical potential. So,  the value of $C_q$ at $V_g\simeq V_g^\text{CNP}$ should increase with growing $B$, achieve the maximal value  at $B=B_c$ and, then, decrease with the further increase of magnetic field even when the Landau levels are rather broadened. When  the Fermi level is located in the bands at $V_g \neq V_g^\text{CNP}$, the maximum in the dependence $C_q(B)$ should be also observed  but in the higher magnetic fields as compared with $B_c$.

In the structures with normal energy spectrum, the anomalous Landau levels are moving apart and the density of states in the gap (which is nonzero due to smearing) decreases with growing magnetic field. Thus, the capacitance should decrease with $B$ near CNP.

\begin{table}
\caption{The parameters of  heterostructures under study}
\label{tab1}
\begin{ruledtabular}
\begin{tabular}{cccccc}
  \# & $d$ (nm) & $p_{s}$ (cm$^{-2}$)& $B_{c}$~(T) & $Q/e$ (cm$^{-2}$)\footnote{At $V_g=0$.} & $p_1/p_2$\\
\colrule
  1122 & 5.6    & $1.3\times 10^{11}$    & $-$ & $2.5\times 10^{11}$   &   $1.8-2.2$ \\
H724 & 5.8    & $1.5\times 10^{11}$   & $-$ & $1.0\times 10^{11}$   &  $2.0-2.2$  \\
1123 & 6.0    & $6.0\times 10^{10}$   & $-$ & $1.5\times 10^{11}$   &    $2.0-2.3$\\
1121 & 6.3    & $8.0\times 10^{10}$   & $0.6\pm 0.2$ & $7.6\times 10^{10}$   &    $2.2-2.5$\\
1023 & 6.5    & $3.6\times 10^{10}$   & $1.1\pm 0.2$ & $1.0\times 10^{10}$   &    $2.3-2.7$\\
     &        &                       &              & $-7.0\times 10^{10}$\footnotemark[2]   &    \\
1022 & 6.7    & $1.4\times 10^{10}$   & $1.2\pm 0.2$ & $1.5\times 10^{10}$   &    $2.3-2.8$ \\
     &        &                       &              & $-1.5\times 10^{10}$\footnotemark[2]   &    \\
1124 & 7.1    & $3.4\times 10^{10}$   & $\sim 0$ & $2.2\times 10^{11}$   & $2.2-2.8$
\end{tabular}
\end{ruledtabular}
\footnotetext[2]{After illumination.}
\end{table}

\begin{figure}
\includegraphics[width=\linewidth,clip=true]{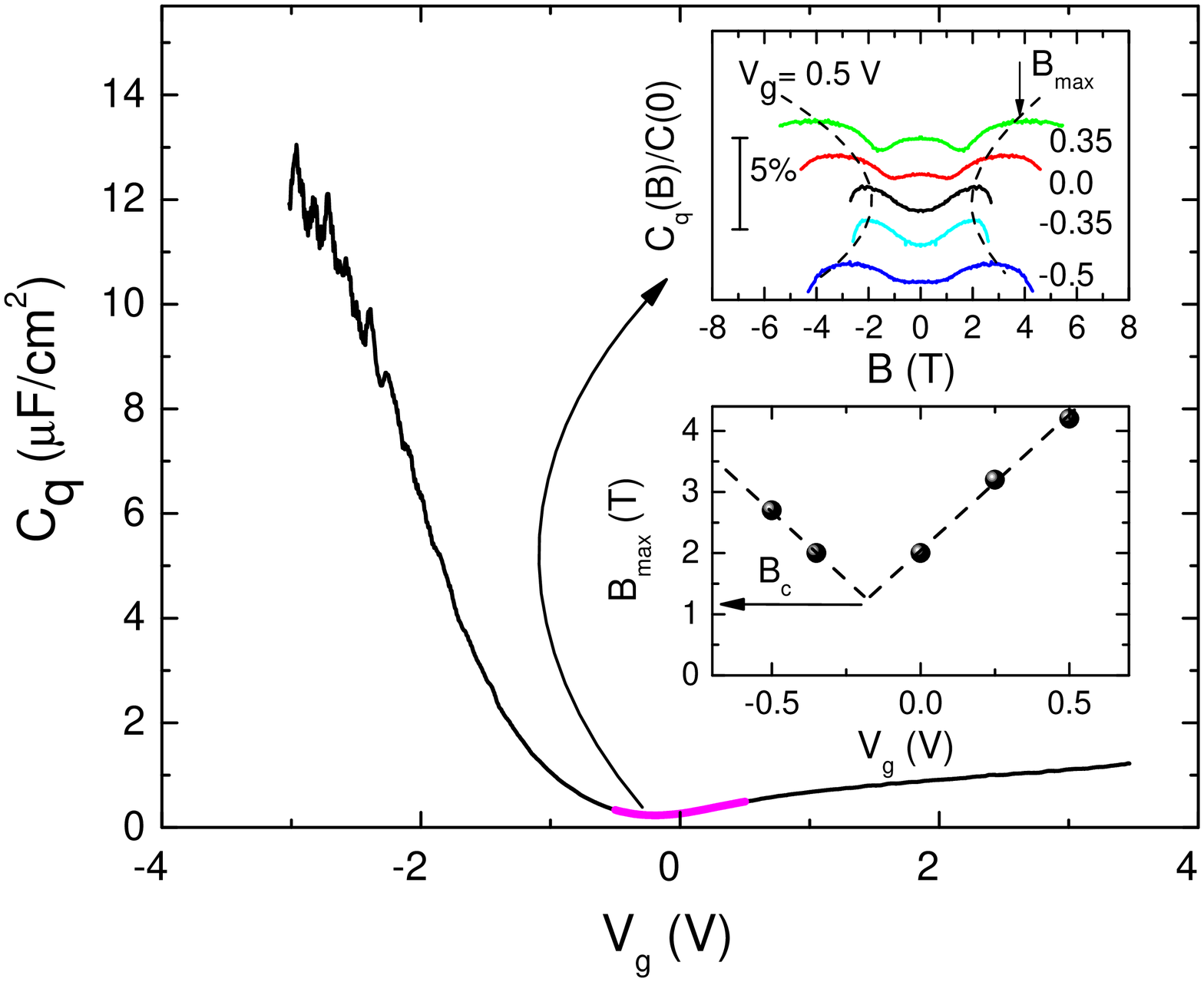}
\caption{(Color online) The gate voltage dependence of the quantum capacitance $C_q$ for the  structure 1023 at $T=4.2$~K and  $B=0$. The upper inset is the magnetic field dependences of  $C_q$ at the gate voltages close to CNP. The lower inset shows the $B_{max}$ values plotted against the gate voltage. }\label{f2}
\end{figure}

The measurements of the quantum capacitance were taken in all the structures investigated. In Fig.~\ref{f2} we have presented, as an example, the  dependence $C_q(V_g)$  measured at $B=0$ on the structure 1023. The nonmonotonic  volt-capacitance  characteristic is consequence of the nonmonotonic energy dependence of the density of states. The minimum in the vicinity of CNP is observed when the Fermi level goes through the energy gap where the density of states is much smaller than that in the valence and conduction bands \footnote{The details of capacitance measurements for HgTe quantum wells of the different widths are beyond the scope of this paper and will be published elsewhere.}. The upper inset in Fig.~\ref{f2} shows the magnetic field dependences of $C_q$ at different $V_g$ near CNP. The maximum in these curves at $B=B_{max}$, which shifts to the higher $B$ values when $V_g$  deviates from CNP, is clearly evident. As seen from the lower inset in Fig.~\ref{f2} the $B_{max}$ vs $V_g$ plot has a minimum and just the minimal value of $B_{max}\simeq 1.2$~T corresponds to the magnetic field $B_c$ in which the anomalous Landau levels $0$ and $-2$ cross each other. The analogous maximum in the dependence $C_q(B)$ at $V_g$ close to $V_g^\text{CNP}$ was also observed in the structures 1022 and 1121. The  $B_c$ values obtained in such a manner for these three  structures are listed in the Table~\ref{tab1}.

As for the remaining four heterostructures, the quantum capacitance for the structure 1124 in the vicinity of CNP is practically independent  on $B$, while for the structures 1122, H724, and 1123  $C_q$ decreases when $B$ increases.

Thus the results of capacitance measurements in the presence of magnetic field show unambiguously  that the structures 1022, 1023 and 1121 has inverted spectrum, the structure 1124 is very close to the critical point $d=d_c$, while the other structures have normal spectra.

\section{The valence band spectrum}
\label{sec:vbs}

\begin{figure}
\includegraphics[width=\linewidth,clip=true]{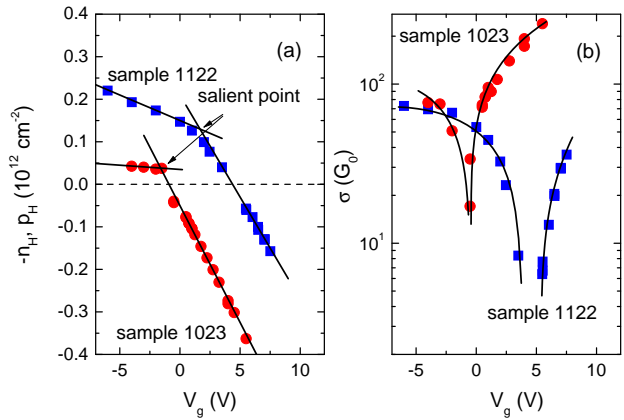}
\caption{(Color online) The gate voltage dependences of (a) the Hall carrier densities $n_\text{H}$ and $p_\text{H}$ and (b) the conductivity at $B=0$ for the  structures  1122 and 1023 with $d<d_c$ and $d>d_c$, respectively. $T=4.2$~K. }\label{f3}
\end{figure}

Let us now inspect the gate voltage dependences of the Hall carrier density obtained at low magnetic field [where $R_H(B)$ is constant] as follows: $n_\text{H}=-1/eR_H(0.1\text{~T})$ when $R_H<0$ for electron density and  $p_\text{H}=1/eR_H(0.1\text{~T})$ when $R_H>0$ for hole density. These dependences are plotted in Fig.~\ref{f3}(a) for structures 1122 and 1023 with normal and inverted spectrum, respectively. The corresponding $V_g$ dependences of the conductivity  are shown in Fig.~\ref{f3}(b). These structures are chosen as typical ones, the results for other structures are analogous. In what follows we will demonstrate all the results namely  for these structures excepting the cases when the simultaneous analysis of results for all structures is useful. One can see that the electron density  depends linearly on $V_g$ over the whole gate voltage range. Note that the slope $dn_\text{H}/dV_g$ coincides  with the value $C/eS$  (where $C$ is the capacitance between the 2D gas and gate electrode and $S$ is the gate area), to within experimental error. The hole density  changes  with the same slope, $dp_\text{H}/dV_g=-dn_\text{H}/dV_g$, within a restricted $V_g$ range only. At the gate voltage $V_g=V_g^\text{s}$, where the hole density achieves some value  $p=p_s$, a salient point is observed. Analogous dependence $p(V_g)$ was observed earlier in Refs.~\cite{Kozlov12, Minkov15}. As seen the $p_s$ values are significantly  different for these structures:  $p_s\simeq 1.3\times 10^{11}$~cm$^{-2}$ for structure 1122 and $p_s\simeq 0.35\times 10^{11}$~cm$^{-2}$ for  structure 1023. The decrease of $|dp_\text{H}/dV_g|$ below $V_g^\text{s}$ can result from  appearance of large enough density of states at the Fermi level, which pins it. They can be the states in the secondary maxima of the valence band spectrum at $k\neq 0$ [see Fig.~\ref{f1}, panels (b), (d), and (e)]  or some states in the barriers or at the interfaces between the layers forming the quantum well.

\begin{figure}
\includegraphics[width=\linewidth,clip=true]{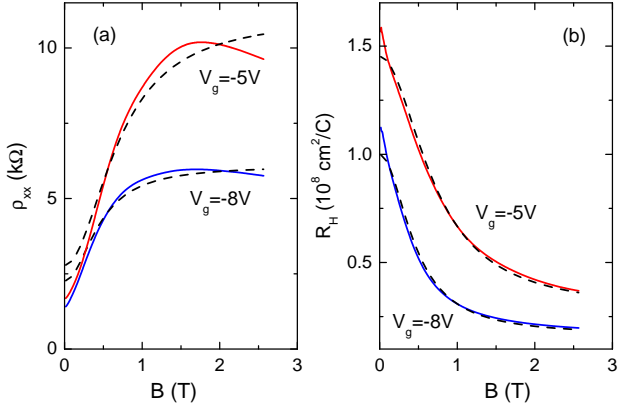}
\caption{(Color online) The  magnetic field dependences of $\rho_{xx}$ (a) and $R_H$ (b) for two gate voltages  for the structure 1023, $T=18$~K. The solid curves are measured experimentally, the dashed ones are the results of the best fit with the parameters shown in Fig.~\ref{f5}.}\label{f4}
\end{figure}

\begin{figure}
\includegraphics[width=\linewidth,clip=true]{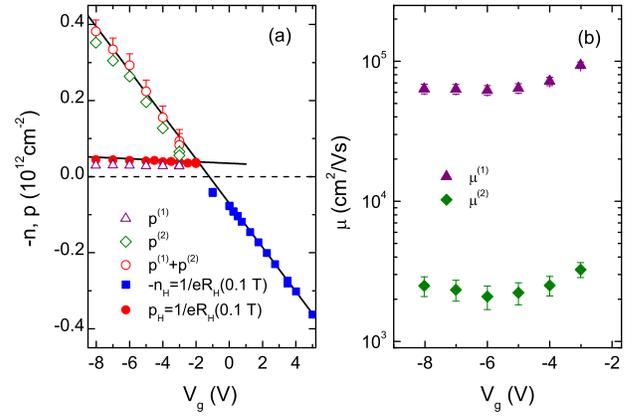}
\caption{(Color online) The gate voltage dependences of the carrier densities (a) and mobilities (b) found from the fit of the magnetic field dependences of $\rho_{xx}$ and $R_H$ shown in Fig.~\ref{f4}. }\label{f5}
\end{figure}

To understand which of these reasons is the primary one we analyze the magnetic field dependences of $R_H$ and $\rho_{xx}$ in classically strong magnetic fields at relatively high temperature at which  the SdH oscillations are suppressed. As an example, in Fig.~\ref{f4} we have presented the $R_H$~vs~$B$ and $\rho_{xx}$~vs~$B$ plots measured on the structure 1023 at $T=18$~K for two gate voltages. These dependences are typical for two-type carries conductivity. The simultaneous fitting of these plots by the standard hand-book expression \cite{Blatt} with the use of hole densities $p^{(1)}$ and $p^{(2)}$, and hole mobilities $\mu^{(1)}$ and $\mu^{(2)}$ as the fitting parameter gives reasonable agreement with the data (see Fig.~\ref{f4}). The hole densities $p^{(1)}$ and $p^{(2)}$ found by this way and the total density $p^{(1)}+p^{(2)}$  together with the hole Hall density $p_\text{H}=1/eR_H(0.1\text{~T})$ for different $V_g$ values are plotted in Fig.~\ref{f5}(a). It is seen that $p^{(1)}$ is close to $p_\text{H}$ and the $p^{(1)}+p^{(2)}$ points fall on the straight line which describes the $n_\text{H}$~vs~$V_g$ data.  The values of the mobility of the second type holes are about $(2-3)\times 10^3$~cm$^2$/V~s [see Fig.~\ref{f5}(b)], which beyond doubt correspond to a free carrier conductivity.

\begin{figure}
\includegraphics[width=0.8\linewidth,clip=true]{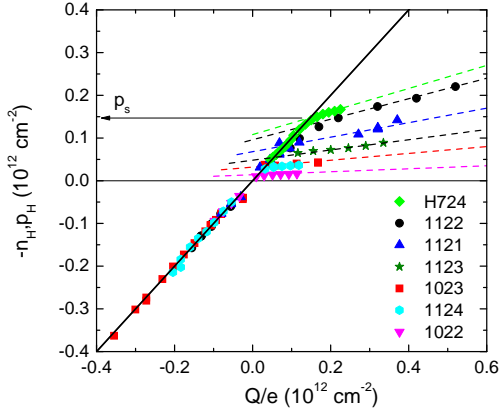}
\caption{(Color online) The Hall carrier density plotted against the charge in the quantum well. As an example arrow shows how the $p_s$ value is found for sample H724. }\label{f6}
\end{figure}

\begin{figure}
\includegraphics[width=\linewidth,clip=true]{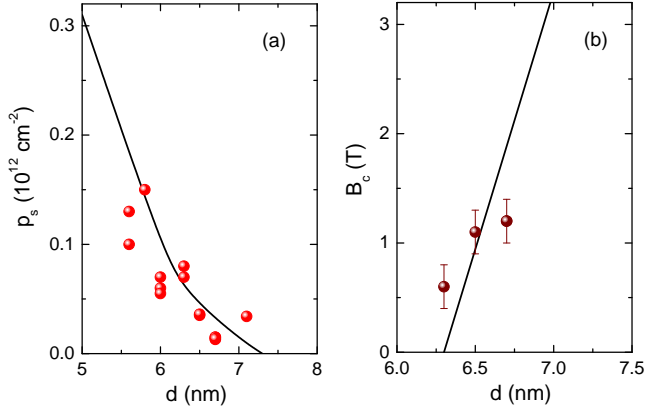}
\caption{(Color online) The values of $p_s$ (a) and $B_c$ (b) plotted against the nominal quantum well width. Symbols are the data, the lines are theoretical results \cite{ZholudevPhD}.}\label{f7}
\end{figure}

Thus, one can argue that the decrease of $|dp_\text{H}/dV_g|$ at $V_g<V_g^\text{s}$ is result of occupation of the secondary maxima of the valence band when the Fermi level approaches these maxima with the decreasing gate voltage. For this case  the value of  $p_s$ should monotonically depend  on the quantum well width. To compare the data obtained for the different structures with different carrier density at $V_g=0$, different insulator thickness and, hence, the different $dn/dV_g$ value we have plotted  the values of the Hall carrier densities  $1/eR_H(0.1\text{~T})$ against the charge in the quantum well $Q=C(V_g-V_g^\text{CNP})$ for all the structures investigated in Fig.~\ref{f6}.
The values $p_s$ found from  Fig.~\ref{f6} are plotted against the nominal quantum well width in Fig.~\ref{f7}(a). The corresponding theoretical dependence calculated within framework of the $8\times 8$ Kane model in Ref.~\cite{ZholudevPhD} is presented also. Taking into account the uncertainty in the real quantum well width and error in $p_s$ determination it can be concluded that the value of $p_s$ in such a type of heterostructures can be used to estimate the quantum well width. Note, the experimental values of $B_c$ are also in a satisfactory agreement with calculation results as evident from Fig.~\ref{f7}(b).

Before closing this section let us discuss the specific feature in the behavior of the hole density evident at $p>p_s$. As seen from Fig.~\ref{f6}, the higher is the $p_s$ value, the larger is the slope of the dependence $p_\text{H}(Q)$ at $p_\text{H}>p_s$.  Since the Hall coefficient $R_H(0.1\text{~T})$ gives the density  of holes only in the central maximum, the slope of the dependence $p_\text{H}(Q)$ at $p_\text{H}>p_s$ is determined by the relation between the densities of hole states in the central and secondary maxima  $\nu_1$ and $\nu_2$, respectively: $e\,dp_\text{H}/dQ=1/(1+\nu_2/\nu_1)$. Thus, the increase of $dp_\text{H}/dQ$ with growing $p_s$ is in qualitative agreement with the fact that the density of states in the central maximum $\nu_1$ at the energy of secondary maximum increases with the energy increase due to the nonparabolicity of dispersion law $E(k)$.

\section{Spin-orbit splitting of valence and conduction bands}
Now let us turn to a more detailed study of the  band spectra. To do this we have measured the SdH oscillations and their angle dependence over the wide carrier density range.

\subsection{Normal band ordering}
\label{ssec:nbo}

\subsubsection{Valence band}
\label{sssec:nbovb}

\begin{figure}
\includegraphics[width=\linewidth,clip=true]{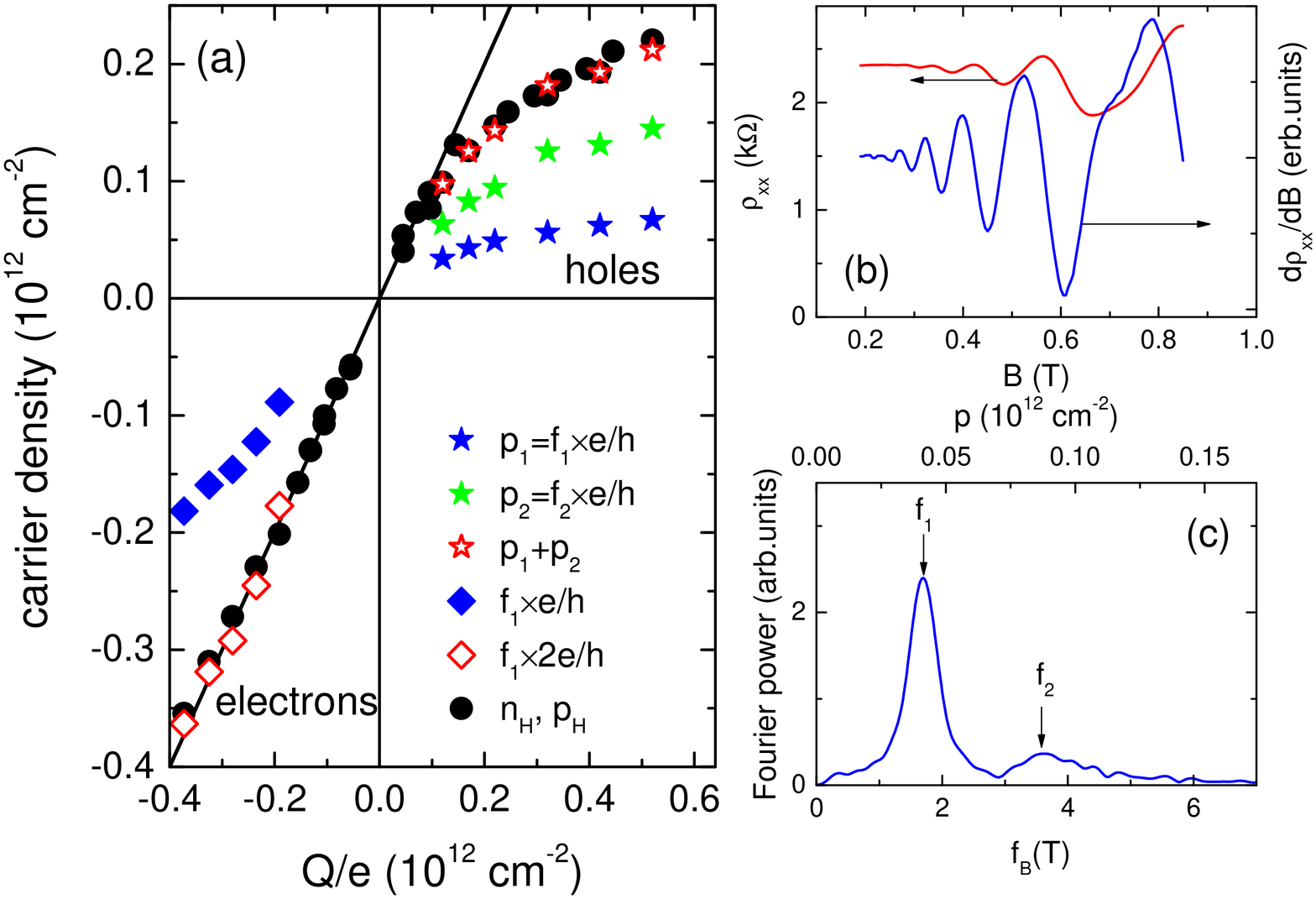}
\caption{(Color online) (a) The carrier density plotted against the charge in the quantum well. The circles are the Hall carrier density, the stars and diamonds are the carrier densities obtained from analysis of the SdH oscillations as described in the text. (b) $\rho_{xx}$ and $d\rho_{xx}/dB$ plotted as a function of magnetic field for $Q/e=1.3\times 10^{11}$~cm$^{-2}$. (c) The result of the Fourier transformation of the SdH oscillations shown in the panel (b). Structure 1122.}\label{f8}
\end{figure}

Qualitatively the results obtained for structures 1122  and their interpretation are analogous to those published in Ref.~\cite{Minkov14}. Nonetheless, we briefly describe the key results that are important for the interpretation of results  both for valence and conduction bands for structures with normal and inverted spectra. The gate voltage dependence of the Hall density, the dependences $\rho_{xx}(B)$ and $d\rho_{xx}/dB$ taken for the structure 1122 at $Q/e=1.3\times 10^{11}$~cm$^{-2}$, and the corresponding Fourier spectrum of the SdH oscillations are presented in Fig.~\ref{f8}.

As seen from Fig.~\ref{f8}(c), two maxima with the frequencies $f_1$ and $f_2$ are easily detected in the Fourier spectrum.
Noteworthy  is that the ratio of the frequencies is close to $2$, i.e., the Fourier spectrum is similar to the case when the spin splitting of the  Landau levels  manifests itself with the magnetic field increase. In such a situation the carrier density should be determined as  $p_\text{SdH}=f_1\times 2e/h$. So, we obtain
$p_\text{SdH}=(0.88\pm 0.05)\times 10^{11}$cm$^{-2}$ for this concrete case. However, the Hall density at this $Q$ value is significantly larger, $p_\text{H}=1.31\times 10^{11}$~cm$^{-2}$ as seen from Fig.~\ref{f8}(a).

Such a difference between the hole density found from the SdH oscillations within the proposed model and the hole density found from the Hall effect takes place over the whole hole density range in all the structures investigated. Thus the interpretation described above does not correspond to our case.

The only interpretation that adequately describes our results is as follows. Each peak in the  Fourier spectrum corresponds to the subband H1 which is strongly split by SO  interaction  into two subbands  H1+ and  H1-. In this case the ``spin''  degeneracy is lifted, and, hence, the hole densities should be found as $p_{1,2}=f_{1,2}\times e/h$, where indexes $1$ and  $2$ correspond to H1+ and H1- subbands.  So, the total hole density is $p_\text{SdH}=p_1+p_2=(f_1+f_2)\,e/h$. The results of such a data interpretation are presented in Fig.~\ref{f8}(a) within whole hole density range by stars. One can see that $p_\text{SdH}$ coincides  with $p_\text{H}$ within experimental error.
Thus, we conclude that the valence band is strongly split by SO interaction so that the ratio of hole densities in the split subbands is about $2$ over the whole hole density range.

\subsubsection{Conduction band}
\label{sssec:nbocb}

\begin{figure}
\includegraphics[width=\linewidth,clip=true]{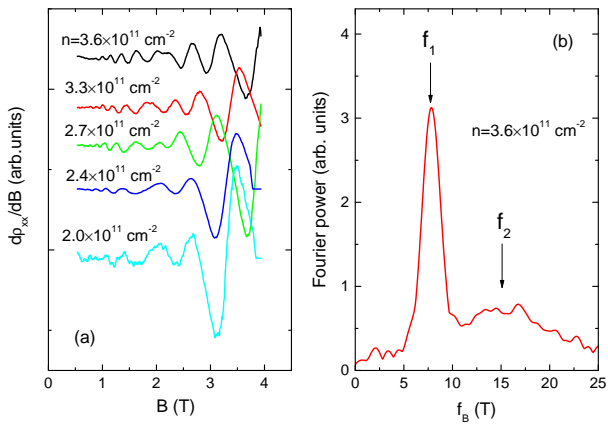}
\caption{(Color online)  The SdH oscillations of electron conductivity measured for structure 1122 at different electron densities (a) and the results  of  Fourier transformation of SdH oscillations for $n=3.6 \times 10^{11}$cm$^{-2}$ (b).}\label{f9}
\end{figure}

To obtain the information on the energy spectrum and splitting of the conduction band we have thoroughly studied the electron SdH oscillations \footnote{Due to large noisiness at positive gate voltage we could not do this in the previous our paper \cite{Minkov14}. The possible reason of the noise  can be large resistance of the \emph{p-n} junctions which are formed in the 2D gas under the gate electrode edge because these structures are of \emph{p} type at $V_g=0$.}.
In Fig.~\ref{f9}(a) we have presented the SdH oscillations for some Hall densities and in Fig.~\ref{f9}(b), as example, the Fourier spectrum for one of them. As seen from Fig.~\ref{f9}(b)  the Fourier spectrum consists of two peaks which characteristic frequencies differ by a factor of two analogously to the case  of the hole conductivity  [see Fig.~\ref{f8}(c)].
The main contribution to the peak with the frequency $f_1$ comes from the low-field SdH oscillations. Therewith, unlike the oscillations in the hole domain, the electron density found  as  $n_\text{SdH}=f_1\times 2e/h$  is very close to the Hall density as evident from Fig.~\ref{f8}(a) indicating that the Zeeman splitting is not resolved. The peak  $f_2$  comes from the oscillations in the high-magnetic fields and it originates from the spin split Landau levels.  This corresponds to the case when the SO splitting of the conduction band is small enough.

Thus, the analysis of SdH oscillations carried out over the whole carrier density range gives surprising result: the valence band is strongly split by SO interaction, while the splitting of the conduction band in the same structure does not reveal itself.

\subsubsection{The Shubnikov-de Haas effect in tilted magnetic field}
\label{sssec:tilted}

To ensure that our interpretation is correct one can explore the SdH oscillations in tilted magnetic field. Really, when the SO splitting is significantly smaller than the Zeeman energy ($\Delta_{SO}\ll \textsl{g}\mu_B B$), the energy of orbital quantization depends on the normal component of the magnetic field $B_\perp=B\,\cos{\theta}$, where $\theta$ is the angle between the magnetic field direction and the normal to the 2D gas plane, while the energy of spin splitting depends on total $B$. At the low magnetic field, where the SdH oscillations are unsplit, this should manifest itself  as strong angle dependence of the  oscillation amplitude. In opposite case of strong spin-orbit interaction ($\Delta_{SO}\gg \textsl{g}\mu_B B$), the spin is rigidly coupled with the orbital motion and the SdH oscillations should be determined  by the normal component of the magnetic  field only.

\begin{figure}
\includegraphics[width=\linewidth,clip=true]{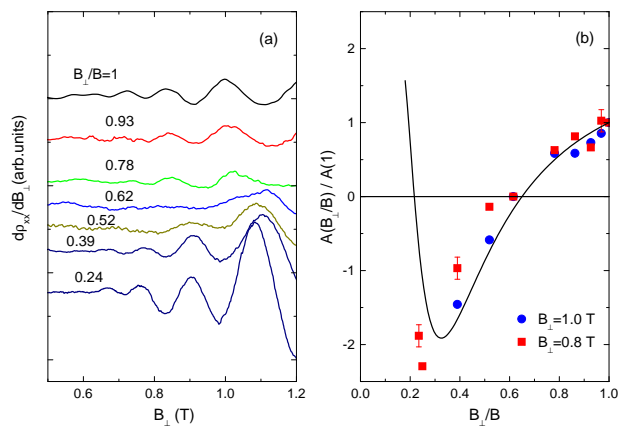}
\caption{(Color online)  The SdH oscillations of electron conductivity  for structure 1122 at different $B_\perp/B$ values (a) and the oscillation amplitude plotted against the $B_\perp$ to $B$ ratio for $B=0.8$~T and $1.0$~T (b), $n=2.5\times 10^{11}$~cm$^{-2}$. Solid line in (b) is the dependence Eq.~(\ref{eq10}) with $X=0.32$. The inversion of the amplitude sign corresponds to  the change of the oscillation phase on $\pi$. }\label{f10}
\end{figure}

The oscillations of $d\rho_{xx}/dB_\perp$ at $B_\perp<1.5$~T for electron density $2.5\times 10^{11}$~cm$^{-2}$ for several tilt  angles are presented in Fig.~\ref{f10}(a). The oscillations within this magnetic field range are caused by  unsplit Landau levels. It is clearly seen that the oscillations decrease in the amplitude when $B$ deviates from the normal orientation, practically disappear at $B_\perp/B\simeq 0.6$, then change the phase and increase in the amplitude at further decrease of the $B_\perp$ to $B$ ratio. This behavior  results from the change of the ratio between the spin and cyclotron energies  $X=\textsl{g} \mu_B B/\hbar\omega_c$  with tilt angle. Supposing  $\textsl{g} \mu_B B$ and $h\omega_c$ are proportional to the total magnetic field  and normal component of $B$, respectively, one can obtain the following expression for the angle dependence of the oscillation amplitude
\begin{equation}
\frac{A(B_\perp/B)}{A(1)} = \cos{\left(\pi X \frac{B}{B_\perp}\right)}/
\cos{\left(\pi X \right)}.
\label{eq10}
\end{equation}
This dependence together with experimental data is shown in Fig.~\ref{f10}(b). One can see that a good agreement is observed when $X=0.32$.

\begin{figure}
\includegraphics[width=0.8\linewidth,clip=true]{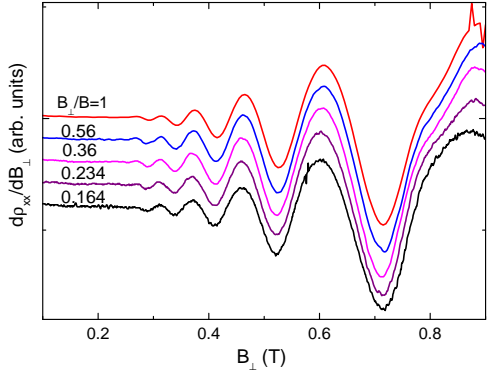}
\caption{(Color online)  The SdH oscillations of the hole conductivity plotted against the normal component of magnetic field  for structure H1122 at different angles, $p=1.35\times 10^{11}$~cm$^{-2}$. }\label{f11}
\end{figure}

Let us now  inspect the behaviour of the SdH oscillations with the changing tilt angle  measured on the same heterostructure in the hole domain (Fig.~\ref{f11}). One can see that unlike the case of the electron conductivity, the SdH oscillations plotted against the normal component of the magnetic field remain practically unchanged with the changing tilt angle. This means that they  are determined by the normal components of magnetic field only. As discussed above such a behavior should be observed when SO interaction is relatively strong: the SO splitting is much greater than the Zeeman energy. Or, alternatively, the independence of the oscillation picture on the tilt angle may caused as well by a strong anisotropy of the effective $\textsl{g}$-factor,  when the in-plane $\textsl{g}$-factor is much smaller than the perpendicular one, $\textsl{g}_\parallel\ll \textsl{g}_\perp$. However, the estimation made with the use of Eq.~(35) from Ref.~\cite{Raichev12-1} shows that it is not the case, the Zeeman splitting  differs in the perpendicular and longitudinal orientations of the magnetic field only slightly for the actual quantum well widths and actual Fermi energies.

Thus, the analysis of the angle dependences of SdH oscillations also shows that  the valence band spectrum is strongly split due to spin-orbit interaction, while the conduction band remains unsplit.

\subsection{Inverted band ordering}
\label{ssec:ibo}

\begin{figure}
\includegraphics[width=\linewidth,clip=true]{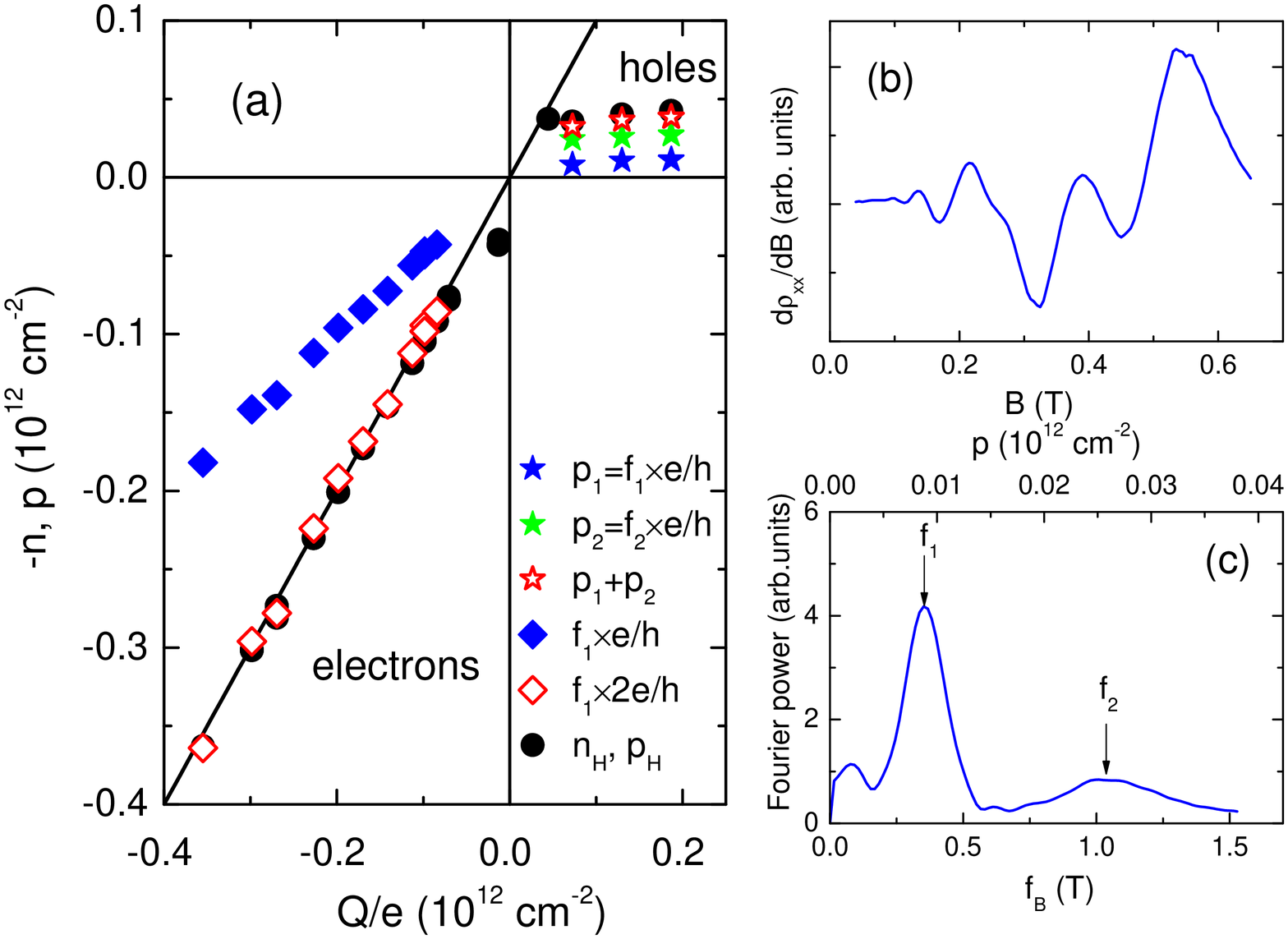}
\caption{(Color online) Results analogous to those presented in Fig.~\ref{f8}, but for the structure 1023 with inverted spectrum.}\label{f12}
\end{figure}

The dependence of carrier densities found from the Hall and Shubnikov-de Haas effects on the charge of 2D gas  for structure 1023 with inverted spectrum, $d=6.5$~nm, is plotted in Fig.~\ref{f12}(a).
In general, it is similar to that for structure with normal spectrum excepting the fact that the $p_s$ value is significantly less in structures with $d>d_c$: namely $p_s\simeq 3.0\times 10^{10}$~cm$^{-2}$  for structure 1023  instead of $\simeq 1\times 10^{11}$~cm$^{-2}$ for structure 1122 with normal spectrum. As discussed in Section~\ref{sec:vbs}  this results from the lesser energy distance between the top of the valence band at $k=0$ and the secondary maxima  at $k\neq 0$.

The SdH oscillations in the hole conductivity regime and their Fourier spectrum are shown in  Fig.~\ref{f12}(b) and Fig.~\ref{f12}(c), respectively. As in the structures with normal spectrum, the only way, which allows us to reconcile the Hall density and data obtained from the SdH oscillations is to assume that each maximum in the Fourier spectrum is associated with the split subband.

\begin{figure}
\includegraphics[width=\linewidth,clip=true]{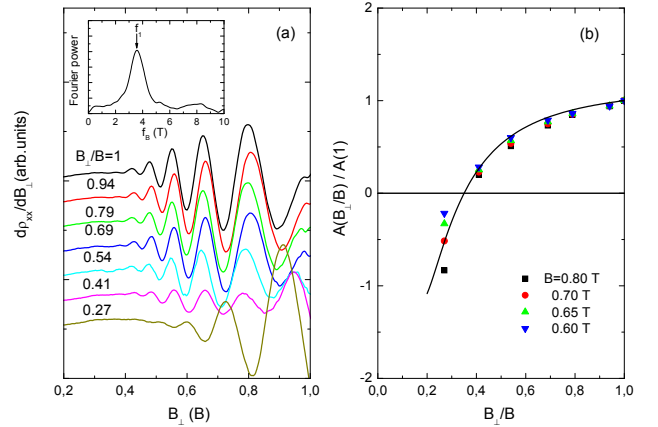}
\caption{(Color online)  The SdH oscillations of electron conductivity  measured for structure 1023 at different angles (a) and the oscillation amplitude plotted against the $B_\perp$ to $B$ ratio for four $B$ values (b), $n=1.7\times 10^{11}$~cm$^{-2}$. Solid line in (b) is the dependence Eq.~(\ref{eq10}) with $X=0.18$. The inset in (a) is the Fourier spectrum for $B_\perp/B=1$.}\label{f13}
\end{figure}

In the electron conductivity regime, the SdH oscillations measured at different tilt angles  are presented in Fig.~\ref{f13}(a) for the Hall density $n_H=1.7\times 10^{11}$~cm$^{-2}$. The Fourier spectrum of the oscillations at normal $B$ orientation is shown in the inset of Fig.~\ref{f13}(a).

The charge dependences of the electron densities found as  $f_1\times 2e/h$ and  $f_1\times e/h$ are shown in Fig.~\ref{f12}(a).  It is seen that  $n_\text{SdH}=f_ 1\times 2e/h$  coincides with $n_H$. It  means that each peak of the SdH oscillations corresponds to two fold degenerated Landau level.  As Fig.~\ref{f12}(a) shows, such a coincidence is observed over the whole range of the electron density. Thus, the spin-orbit splitting of the conduction band in the structures with the inverted spectrum does not reveal itself as well as in the structures with normal spectrum.

The conclusion about strong  splitting of the valence band and weak splitting of  the conduction band is consistent with the behavior of the oscillations in tilted magnetic field.

The behavior of the oscillations of the electron conductivity with the tilt angle shown in Fig.~\ref{f13}(a) is analogous to that presented in Fig.~\ref{f10}(a) for structure with normal spectrum. The angle dependence of  the oscillation amplitude in Fig.~\ref{f13}(b) demonstrates that the Zeeman splitting becomes equal to half orbital one at $B_\perp/B$ close to $0.35$. The dependence calculated from Eq.~(\ref{eq10}) with $X=0.18$ well describes the data \footnote{The detail analysis of electron-density dependence of $\textsl{g}$-factor in the structures with different width of quantum well are beyond the scope of this study and will be discuss in another paper.}.

\begin{figure}
\includegraphics[width=0.8\linewidth,clip=true]{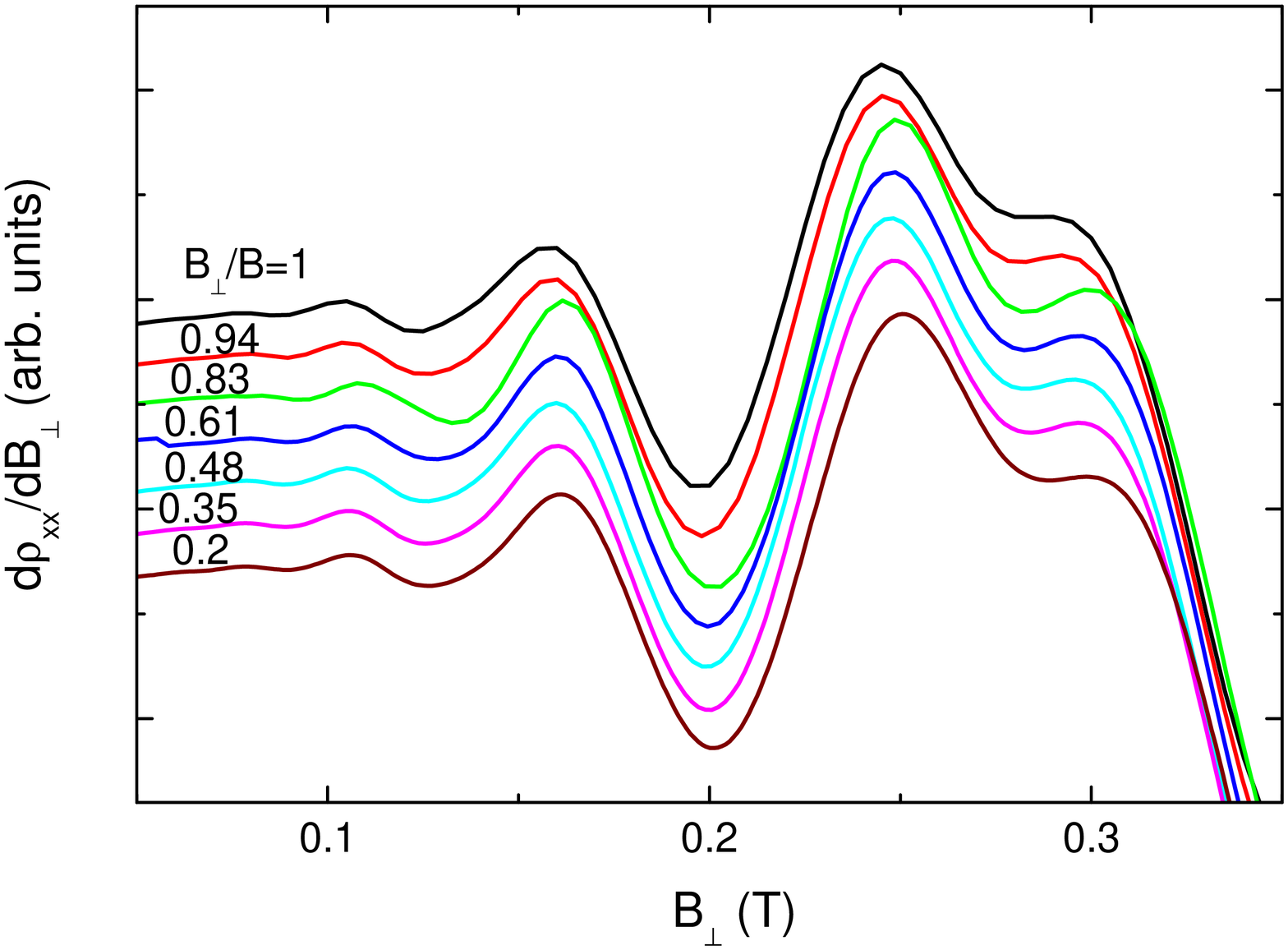}
\caption{(Color online)  The SdH oscillations of hole conductivity measured at different angles for structure H1023 with inverted spectrum, $p=3.8\times 10^{10}$~cm$^{-2}$. }\label{f14}
\end{figure}

The SdH oscillations of the hole conductivity measured at different tilt angle are presented  in Fig.~\ref{f14}. As well as for the structures with normal spectrum (see  Fig.~\ref{f11}), neither positions nor amplitudes depend on the tilt angle that corresponds to strong SO  splitting of the valence band (see discussion above).

\begin{figure}
\includegraphics[width=\linewidth,clip=true]{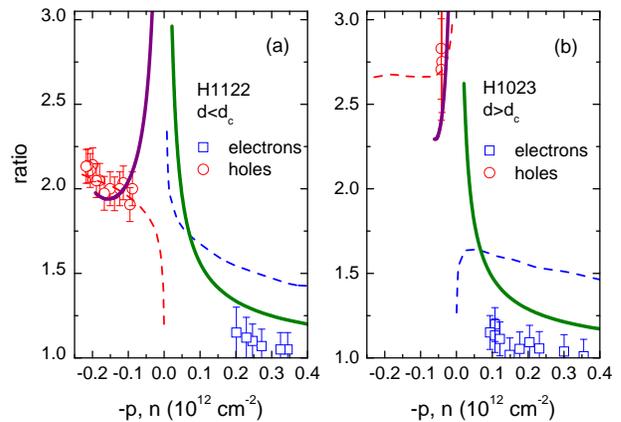}
\caption{(Color online)  The ratio between carrier densities in split subbands plotted against the carrier density for structure 1122 (a) and 1023 (b) with normal and inverted spectra, respectively. Symbols are the experimental results, the dashed line are the result of the \emph{kP} calculations described in the text, the solid lines  are the results obtained by the $sp^3$ tight-binding method \cite{Tarasenko15}. }\label{f15}
\end{figure}

Summing up all the data given above we can formulate the main outcome.
The analysis of the Shubnikov-de Haas and Hall effects over the whole carrier density range in the structures with $d<d_c$ and $d>d_c$ gives surprising result:  the valence band is strongly split by SO interaction in the structures both  with normal and inverted spectrum, while the splitting of the conduction band in the same structures does not reveal itself. To illustrate the written we have plotted  the ratio between the carrier densities in the split subbands in the valence and conduction bands as a function of carrier density in Fig.~\ref{f15} \footnote{We have plotted the ratio $p_2/p_1$ but not the value of SO splitting, $\Delta_{SO}$, because to calculate  $\Delta_{SO}$ from $p_2/p_1$ one needs the carriers effective mass which is known with some error}. This figure is the key result of the paper.

\section{Comparison with results of $k\cdot P$ and tight-binding calculations}
\label{sec:comp}

The strong SO splitting of the valence band at $d<d_c$ was observed in our previous paper~\cite{Minkov14}. The experimental results are interpreted  under assumption that the quantum well is located in the strong electric field of a \emph{p-n} junction that in its turn results in strong Bychkov-Rashba effect. It has been shown that the \emph{kP} model describes the valence band spectrum quantitatively. Unfortunately,  we were  unable in that paper to study the conduction band with the accuracy needed to obtain the SO splitting reliably. For heterostructures studied in the present paper it was possible.

Let us now  compare the experimental results obtained in the present paper both for the valence and conduction bands with the results calculated within framework of the \emph{kP} model used  in Ref.~\cite{Minkov14}.  As before, we use here the six-band Kane Hamiltonian. The direct integration technique is applied to solve the Shr\"{o}dinger equation as described in Ref.~\cite{Lar97}. The other
parameters were the same as in Refs.~\cite{Zhang01,Novik05}. As in the previous paper \cite{Minkov14}, the value of  electric field serves as the fitting parameter, which provides the ratio between the hole densities in the split subbands, which is  observed experimentally (see Fig.~\ref{f15}). It is approximately  equal to $80$~mV$/d$ for all the structures under study.

\begin{figure}
\includegraphics[width=\linewidth,clip=true]{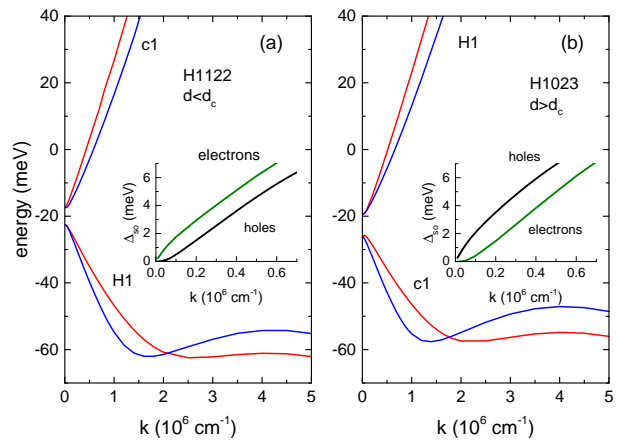}
\caption{(Color online)  The dispersion $E(k)$ for structures 1122 (a) and 1023 (b) with normal and inverted spectra, respectively, calculated within the six-band Kane model in the presence of electric field $80$~mV$/d$. In the inserts, the SO splitting plotted against the quasimomentum value.}\label{f16}
\end{figure}

Let us now inspect the SO splitting of the conduction band. The calculated energy spectrum for the heterostructures with  $d<d_c$ and $d>d_c$  is shown Fig.~\ref{f16}. The energy splitting $\Delta_{SO}$ as a function of the quasimomentum value is shown in the insets. It is seen that the splitting of the conduction band at low $k$ values, $k<0.1\times 10^6$~cm$^{-1}$,  is much larger than that of the valence band for $d<d_c$. For $d>d_c$, the situation is opposite; the valence band is split much stronger than the conduction band. Such relationships between splittings agree with the known result of symmetry analysis according to which $\Delta_{so}\propto k$ for the c1 band and $\Delta_{so}\propto k^3$ for the H1 band near $k=0$. It should be emphasized that so low quasimomentum values correspond to very low carrier densities  ($\sim 10^9$~cm$^{-2}$ and less) which are inaccessible experimentally. At larger quasimomentum values,  these relationships  are violated.  The splitting values of c1 and H1 bands become close to each other. As seen from Fig.~\ref{f16} the difference in the spin-orbit splitting for the valence and conduction bands does not exceed $30$~\% for both heterostructures for $k_F>0.5\times 10^6$~cm$^{-1}$ that corresponds to the actual carrier density range $n,p> 5\times 10^{10}$~cm$^{-2}$ [see Fig.~\ref{f15}].

In order to compare the results of the \emph{kP} calculations with the experimental data, we have calculated the carrier density in the split subbands as $k^2/(4\pi)$ and plotted the $n_2$ to $n_1$ and $p_2$ to $p_1$ ratios against the total carrier density in Fig.~\ref{f15} by the dashed lines. It is evident that  this model perfectly describes the data relating to the hole split subband and  gives no agreement with the experimental results concerning the conduction band splitting.  One of the reason why the used \emph{kP} model fails when compared with the experiment is the fact that it ignores the asymmetry caused by the difference of the quantum well interfaces.

The role of the bulk inversion asymmetry of host crystals and interface inversion asymmetry in the SO splitting of the energy spectrum of 2D carriers was recently studied in Ref.~\cite{Tarasenko15}. Using the symmetry analysis and atomistic calculations the authors obtain surprising result.  The asymmetry of CdTe/HgTe and HgTe/CdTe interfaces forming the CdTe/HgTe/CdTe quantum well results in the giant splitting of the energy spectrum; in the quantum wells of critical and close-to-critical width, $d\simeq d_c$, the splitting reaches a value of about $15$~meV.

Analogous calculations have been performed for our concrete case of (013) HgTe. The quantum well was supposed symmetric in the sense that no
electric field was applied across the the well. The microscopical strain has been calculated in the atomistic valence force field model \cite{Keating66} and then incorporated in the tight-binding using standard procedure \cite{Jancu98}. The tight-binding parameters used in calculations were obtained using procedure similar to that described in details in the Supplemental Material to Ref.~\cite{Tarasenko15} with the following modification. In Ref.~\cite{Tarasenko15} it was assumed that the change of the ionicity across the interface may be extracted from the atomic levels obtained in \emph{ab initio} calculations. Instead, one may introduce more physical parameter, the change of electrostatic potential on anion across the interface, and fit this parameter to reproduce the expected properties of the interface. Tight-binding parameters in Ref.~\cite{Tarasenko15} correspond to the change of electrostatic potential on Te between the quantum well and barrier equal to $1$~eV, which is in accordance with \emph{ab initio} calculations of HgTe/CdTe heteropair. For the interface between HgTe and Hg$_{1-x}$Cd$_{x}$Te ($x\simeq 0.5$) alloy this parameter should be reduced to $500$~meV which leads to proportional reduction of the interface-induced spin splitting. The carrier densities in the subbands were calculated as $S_k/(2\pi)^2$, where $S_k$ is the area inside the Fermi contour, which is  not a circle due to absence of axial symmetry. The results of calculations are depicted in Fig.~\ref{f15} by the solid lines.
One can see that the atomistic calculations describe the experimental SO splitting of the spectrum reasonably not only for the valence band but for the conduction band also. It is necessary to stress that no fitting parameters have been used in calculations.

\section{Conclusion}

We have studied the energy spectrum of the conduction and valence bands of 2D states in HgTe quantum well by means of magnetotransport measurements. The structures investigated have the width of the quantum wells close to the critical value $d\simeq d_c$ where the spectrum changes from normal (at $d<d_c$) to inverted (at $d>d_c$). Simultaneous analysis of the SdH oscillations and Hall effect over the wide range of the electron and hole densities gives surprising result: the top of the valence band is strongly split by spin-orbit interaction while the splitting of the conduction band is absent, within experimental accuracy. This conclusion is supported by the results obtained in the tilted magnetic fields. The behavior of the SdH oscillations of electron conductivity with the changing tilt angle corresponds to the case when the orbital quantization is determined by normal component of the magnetic field, while the spin splitting is determined by the total field.  The oscillations of the hole conductivity are totally determined by the normal component of the magnetic field indicating the SO interaction wins the Zeeman effect in the actual magnetic field range.   Surprisingly, but such a ratio of the splittings is observed as for structures with normal spectrum ($d<d_c$) so for structures with inverted one ($d>d_c$). These data are inconsistent with the \emph{kP} calculations which take into account the Bychkov-Rashba effect caused by electric field directed across the quantum well. It is shown that the experimental results can be reasonably described within framework of the tight-binding method which properly takes into consideration the interface inversion asymmetry.

\acknowledgements

We are grateful to I.\,V.~Gornyi, S.\,A. Tarasenko,  and O.\,E. Raichev for useful discussions.
The work has been supported in part by the Russian Foundation for Basic
Research (Grants No. 13-02-00322 and No. 15-02-02072) and by Act 211 Government of the Russian Federation, agreement No. 02.A03.21.0006. A.V.G. and O.E.R. gratefully acknowledge financial support
from the Ministry of Education and Science of the Russian
Federation under Projects No. 3.571.2014/K and No. 2457.

\end{document}